# Highly controlled coating of a biomimetic polymer in TiO$_2$ nanotubes

Gabriel Loget [a,1], Jeung Eun Yoo [a,1], Anca Mazare [a], Lei Wang [a], Patrik Schmuki [a,b,*]

[a] Department of Materials Science and Engineering, WW4-LKO, University of Erlangen-Nuremberg, Martensstrasse 7, D-91058 Erlangen, Germany

[b] Department of Chemistry, King Abdulaziz University, Jeddah, Saudi Arabia

[1] These authors contributed equally

[*] Corresponding author. Tel.: +49 9131 85 275 75; fax: +49 9131 85 275 82.

E-mail address: schmuki@ww.uni-erlangen.de (P. Schmuki)

## Abstract

Highly controlled coating of biomimetic polydopamine (PDA) was achieved on titanium dioxide nanotubes (TiO$_2$ NTs) by exposing TiO$_2$ NT arrays to a slightly alkaline dopamine solution. The thin films act as photonic sensitizers (enhancing photocurrents and photodegradation) in the visible light range. The PDA coatings can furthermore be used as a platform for decorating the TiO$_2$ NTs with different co-catalysts and metal nanoparticles (NPs).

# Introduction

Over the past decade, arrays of $TiO_2$ nanotubes ($TiO_2$ NTs) grown on Ti surfaces by self-organizing anodization have attracted a huge scientific interest for applications in the fields of dye-sensitized solar cells (DSSCs), water splitting, photocatalysis and biomedical devices.[1-3] A strong focus lies currently on the decoration and the coating of the $TiO_2$ NTs with elements that can increase the tube performances.[1-2] For instance, organic polymer coatings can be used for sensitizing $TiO_2$ NTs, leading to an improved solar energy conversion[4,5], and decoration with inorganic nanoparticles (NPs) can increase the catalytic or photocatalytic efficiency of the tubes.[6-11]

Polydopamine (PDA) is a synthetic eumelanin polymer mimicking the biopolymer secreted by mussels to attach to surfaces with a high binding strength.[12-14] It exhibits unique adhesive properties and has recently attracted considerable interest as a multifunctional thin film coating. Among others, it has been applied for sensing,[15-18] for the passive dispensing of aqueous solutions,[19] as a sensitizer in DSSCs,[20] for making graphitic transparent stretchable electrodes[21], and it can moreover be used as a platform for the decoration of surfaces with NPs.[22-27]

In this study, we show the feasibility to achieve a highly-controlled coating of the walls of anodized $TiO_2$ NTs with PDA. We demonstrate that PDA sensitizes the arrays and can serve as a platform for the decoration of $TiO_2$ NTs with different co-catalyst materials. We finally use these arrays as photoanodes operating at low overpotential in the visible solar spectrum.

## Experimental Section

HF, $H_3PO_4$, Trizma-base, dopamine hydrochloride, $AgNO_3$, $NaBH_4$, $CoSO_4$, NaOH, methylene blue and $Na_2SO_4$ were purchased from Sigma-Aldrich. Titanium foils were anodized following the recently published procedure.[10,28,29] For PDA coatings, the $TiO_2$ NTs were taped in a beaker containing 30 mL of 10 mM Tris-buffer (pH adjusted to 8.6 with HCl). 120 mg of dopamine were added and the solution was maintained under stirring for a desired time. After coating, the samples were rinsed with mQ water and dried in a nitrogen stream. All the decorations were performed on freshly-coated surfaces. For AgNP-decoration the PDA-coated surface was exposed to a solution of 20 mM $AgNO_3$ in the dark. After 14 h, the surface was rinsed with mQ water and dried in a nitrogen stream. For PtNP-decoration, the PDA-coated surface was taped in a reaction vessel that contained 40 mL of a 0.2 mg.mL$^{-1}$ $H_2PtCl_6$ (Metakem) solution. The solution was maintained at 90°C, and 100 µL of a 50 mM $NaBH_4$ solution was added dropwise under vigorous stirring, which caused darkening of the solution due to the formation of PtNPs. The array was left 30 min in the colloidal solution under stirring, followed by rinsing and drying. For modification with $Co(OH)_x$, the $TiO_2$ NTs were immersed in a mixed solution of 0.1 M $CoSO_4$ and 0.1 M NaOH with a ratio of 1:1 for 15 min, then rinsed and dried in $N_2$. Scanning electron microscope (FE-SEM, Hitachi SEMFE 4800) and X-ray photoelectron spectroscopy (XPS, PHI 5600, US) were used for characterization. Photodegradation was performed under stirring with the $TiO_2$ NT array taped in a quartz cuvette that was filled with an aqueous solution of 0.01 mM methylene blue. A spot of 1.7 cm$^2$ was irradiated with a 473 nm laser (MLB 473, 20 mW) and UV-vis spectra were recorded at the desired times using a spectrophotometer (Perkin Elmer Lambda XLS+). Photoelectrochemical experiments were performed in 0.1 M $Na_2SO_4$ aqueous solution. Photoelectrochemical spectra were measured using a monochromatized Xe lamp. The band gap of PDA was obtained using a Pt surface coated for 3 h with PDA. To evaluate the photoresponse, photocurrent transients were recorded at a constant potential of 0.5 V (vs. Ag/AgCl) with a potentiostat (Jaissle IMP83 PC-T-BC) while the wavelength was varied using a motor driven monochromator (Oriel Cornerstone 130 1/8 m). Water splitting experiments were done with an AM 1.5 solar simulator (300W Xe, Solarlight) equipped with a 400 nm cut-off filter, the potential was controlled by a potentiostat (Jaissle 1002 T-NC12). I-V curves were recorded at 1 mV.s$^{-1}$.

**Results and Discussion**

Fig. 1 shows the highly organized TiO$_2$ NT layers used in this study – the layers had a thickness of ≈200 nm, tube diameters of ≈80 nm, and were fabricated by anodization in HF-containing concentrated phosphoric acid electrolyte.[10,28,29] As illustrated in Fig. 1a, the PDA coatings were prepared by the procedure developed by Lee *et al.*[12], which consists in immersing the array in a solution containing dopamine (in our case, 40 mg.mL$^{-1}$) at a slightly basic pH of 8.6. It is well-known that at this pH, dopamine spontaneously polymerizes, which leads to the coating of the immersed surfaces with PDA.[13,14] For the first proof-of-principle experiment, a rather long immersion time of 24 h was chosen in order to clearly see if PDA can coat the inner cavities of the TiO$_2$ NTs. The scanning electron microscope (SEM) pictures of Fig. 1b and c show cross sections of the tube layer and reveal that PDA coated the walls that are exposed to the solution, even inside the cavities. This shows that the diffusion of dopamine and PDA oligomers is not significantly altered in the tube and that PDA can be reliably used for coating the entire walls of TiO$_2$ NTs.

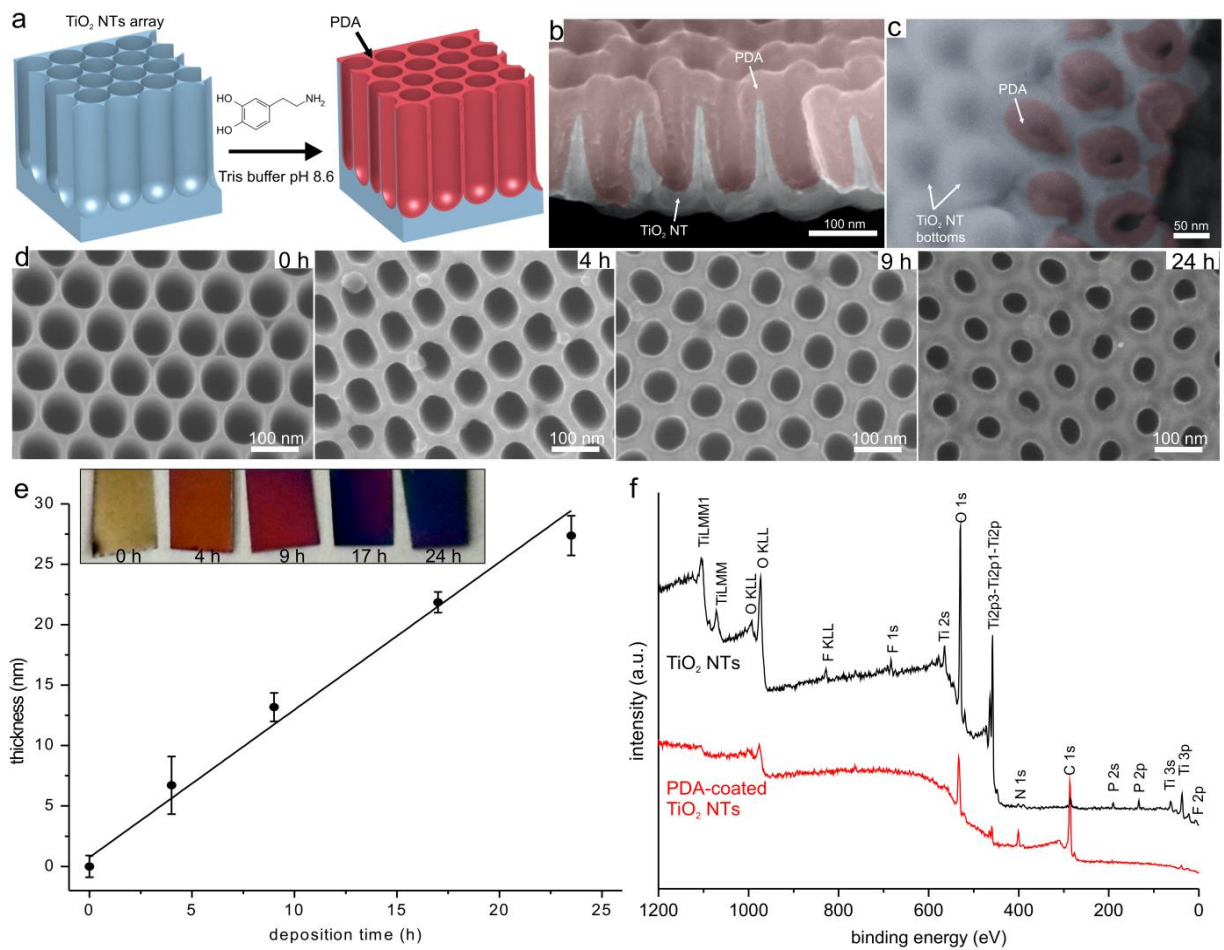

**Fig. 1.** a) Scheme showing the coating of TiO$_2$ NTs with PDA. b, c) Colorized SEM pictures taken after coating and showing cross-sections b) perpendicular and c) parallel to the surface. d) Top view SEM pictures taken at different coating times. e) PDA thickness as a function of the coating time. Inset: Photographs showing the color of the surface at different coating times. f) XPS spectra of uncoated (black) and PDA-coated (red) TiO$_2$ NTs.

The top view of a TiO$_2$ NT array was observed with SEM before coating and at different coating times. The pictures shown in Fig. 1d shows that the diameter of the tubes shrunk with an increasing coating time. The inner TiO$_2$ NT diameters were measured, which allowed determining the PDA thickness as a function of the coating time, shown in Fig. 1e. From this curve, a coating rate of 1.2 nm.h$^{-1}$ was determined. Interestingly, the increasing thickness of PDA changed the color of the surface from yellow to dark blue, as it can be seen in the inset of Figure 1e. XPS measurements of the surfaces were performed before and after 3 h of coating. Fig.1f shows that the bare surface displayed the typical peaks of a bare $TiO_2$ NT surface[1,2,10,28,29] whereas the coated surface revealed a strong decrease of the Ti and the O peaks, a disappearance of the F and P peaks, a drastic increase of the C peak and the appearance of a N peak at 400 eV. Even if the exact molecular structure of PDA is still under debate, it is well known that it contains high amounts of C and N atoms[30] and thus these spectra are in very good agreement with our SEM observations, confirming the presence of an organic polymer layer on the $TiO_2$ NTs. These results prove that highly-controlled PDA coatings can be performed on $TiO_2$ NTs; we will now discuss the photoelectrochemical properties of these PDA coated-$TiO_2$ NT arrays.

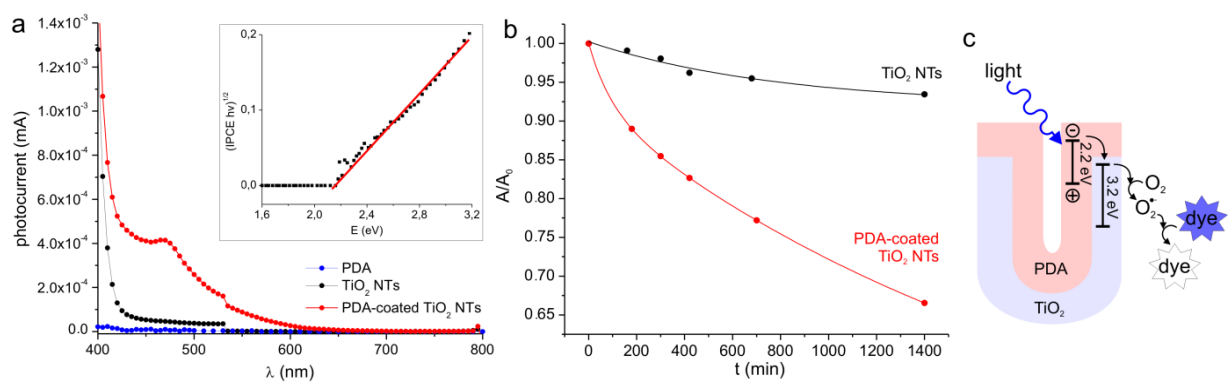

**Fig. 2.** a) Photocurrent spectra. Inset: determination of the band gap of PDA measured with PDA on Pt. b) MB photodegradation by a 473 nm laser. Inset: Scheme showing the photodegradation mechanism.

Due to the relatively low conductivity of PDA layers,[31] we decided to use a rather short coating time of 3 h (which corresponds to a PDA thickness of 3-4 nm) for the rest of this work in order to achieve reasonable current flow. Fig. 2a shows the photocurrent spectra for the bare $TiO_2$ NT, PDA and the PDA-coated $TiO_2$ NT surfaces in the visible spectral range. Clearly, the PDA-coated $TiO_2$ NTs provided an enhanced response, which indicates that PDA can act as sensitizer for the TiO$_2$ NTs. As shown in the inset of Fig. 2a, the measured PDA band-gap was of 2.2 eV, which is very close to the one reported by Nam *et al.*[20]

The PDA-sensitized TiO$_2$ NTs were applied for photodegradation experiments using a blue laser (473 nm, 12.5 mW.cm$^{-2}$) with methylene blue (MB) as a model dye. Fig. 2b shows that the dye degradation was significantly enhanced for the coated surface. This suggests the mechanism depicted in Fig. 2c,[1] where blue light generates an electron-hole pair in the PDA and the electron gets transferred to the TiO$_2$ conduction band for generating superoxide radical which degrades MB.[32,33] Those results demonstrate that PDA thin films can sensitize TiO$_2$ NT arrays; we will now show that in addition PDA can be used for the tube decoration with different materials.

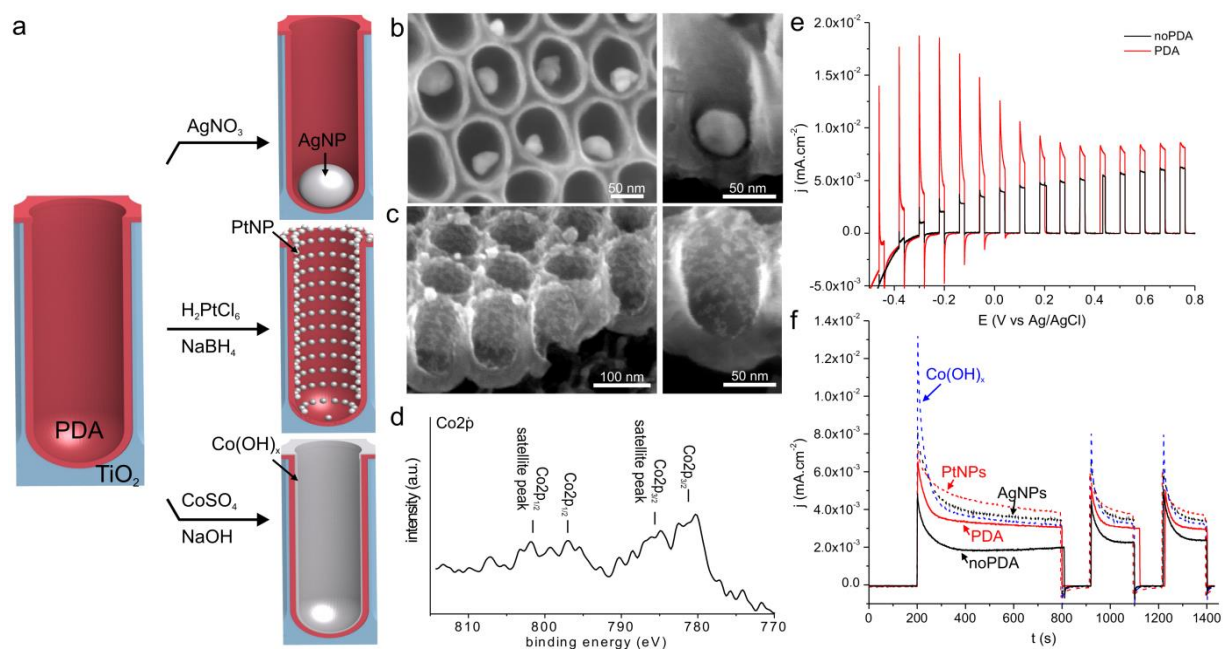

**Fig. 3.** a) Scheme showing the decoration of the PDA-coated tubes with AgNPs, PtNPs and Co(OH)$_x$. b,c) SEM pictures of b) AgNP- and c) PtNP-decorated TiO$_2$ NTs. d) XPS spectrum showing the Co2p peaks. e) Light/dark

---

[1] In this scheme, the HOMO of PDA (-5.25 eV) was placed according to the value reported in Ref[20].

I-V curves showing the influence of PDA coating. f) Transient photocurrents at 0 V vs Ag/AgCl. All the experiments were performed in 0.1 M $Na_2SO_4$ and AM 1.5 illumination condition with a 400 nm cut-off filter.

As described in Fig. 3a, three types of decoration were performed. The first one consisted in using the reductive ability of the PDA catechol groups for the spontaneous reduction of a metal cation on the coated tubes.[24,25] The PDA-coated samples were exposed to a solution of $AgNO_3$ for 14 h in the dark, followed by rinsing and drying under $N_2$. This resulted in the decoration of the tubes with AgNPs having a diameter of 39±9 nm (measurements done on ≈90 NPs), as shown in Fig. 3b. Interestingly, the great majority of the AgNPs (≈80 %) were located inside the tubes cavities with a distribution of one NP per tube (≈95 % of the filled tubes contained only 1 AgNP), which suggests a preferential confined growth of the AgNPs inside the $TiO_2$ NT cavities.

In the second example, we took advantage of the strong adhesive properties of PDA[13,14] for decorating it with PtNPs. In this case, the bulk formation of PtNPs, done by reducing $H_2PtCl_6$ with $NaBH_4$ was performed in presence of the PDA-coated $TiO_2$ NTs and the arrays were kept in the colloidal solution for 30 min, followed by rinsing and drying. Fig. 3c shows that it led to the homogeneous decoration of the tube walls with aggregates of PtNPs having a size of about 1 nm.

In the last example, PDA-coated $TiO_2$ NTs were exposed 15 min to a solution in which the precipitation of $Co(OH)_x$ was induced by mixing $CoSO_4$ and NaOH, a procedure that has been previously applied for the modification $Ta_3N_5$ nanorods with $Co(OH)_x$.[34] In this case, no particles were observed at the SEM but XPS analysis revealed the presence of $Co(OH)_x$ with the Co2P peaks shown in Fig. 4d.[35] From these peaks, a Co amount of 2.3 at% was determined. These decorated arrays were investigated as photoanodes.

Photoelectrochemical water splitting was performed in 0.1 M $Na_2SO_4$, as PDA removal may occur for strongly acidic and alkaline pH.[36,37] Fig. 3e shows the influence of PDA on the photocurrent density under irradiation with the solar visible spectrum. The photocurrents were higher for the coated surfaces with respect to the uncoated ones, with an increase that was more pronounced for low overpotential values. The spike-like peaks that can be observed below 0.3 V for the coated photoanode suggest a higher recombination rate for this electrode. The performances of the coated and decorated $TiO_2$ NTs were compared by potentiostatic experiments performed at 0 V vs Ag/AgCl, shown in Fig. 3f. First, it can be seen that the PDA-decorated sample exhibited a stable photocurrent density of $3.1 \cdot 10^{-3}$ mA.cm$^{-2}$, which represents a photocurrent increase of ≈35% with respect to the uncoated $TiO_2$ NTs. The three decorated photoanodes produced higher photocurrents than the non-decorated ones. $Co(OH)_x$-decorated $TiO_2$ NTs provided the highest photocurrents at first, but showed the strongest decay with time, leading to a photocurrent value almost equal to the one of the non-decorated PDA-coated photoanode. The AgNP-decorated photoanode provided intermediary performances with a photocurrent density of $3.5 \cdot 10^{-3}$ mA.cm$^{-2}$. Contrariwise, the PtNP-decorated photoanode led to the best performances after 50 s of illumination with photocurrents that were ≈70% higher than in the case of uncoated $TiO_2$ NTs. In this case, the increased photocurrents could be explained by an enhanced light absorption[38] and a lower charge recombination rate.[39]

**Conclusion**

We demonstrate a highly controlled coating of $TiO_2$ NTs walls with PDA applying a simple method. The PDA thin films sensitized the arrays and were used for enhancing photodegradation rates and photocurrents in the visible solar spectrum. PDA was furthermore used as a platform for decorating the $TiO_2$ NTs with different materials such as AgNPs, PtNPs, and Co(OH)$_x$. These decorations improved the photoelectrochemical performances of the arrays for water splitting at a low overpotential. Due to the numerous advantages of PDA,[13,14] we are convinced that these findings will be beneficial for the development of new TiO$_2$ NTs-based biomedical devices[40,41] and batteries.[42,43]

## Acknowledgements

This work was supported by a post-doctoral research grant from the Alexander von Humboldt Foundation. ERC and DFG are acknowledged for support.